\documentclass[page-classic]{epl2} 

\title{Particle dynamics in colloidal suspensions above and below the glass-liquid re-entrance transition}
\shorttitle{Particle dynamics in colloidal suspensions ...} 

\author{Andrzej Latka\inst{1} \and Yilong Han\inst{2} \and Ahmed M. Alsayed\inst{3} \and Andrew B. Schofield\inst{4} \and A. G. Yodh\inst{3} \and Piotr Habdas\inst{1}\thanks{E-mail: \email{phabdas@sju.edu}}}
\shortauthor{Andrzej Latka \etal}

\institute{                    
  \inst{1} Department of Physics, Saint Joseph's University - Philadelphia, PA 19131, USA\\
  \inst{2} Department of Physics, Hong Kong University of Science and Technology - Hong Kong\\
  \inst{3} Department of Physics and Astronomy, University of Pennsylvania - Philadelphia, PA 19104, USA\\
  \inst{4} The School of Physics and Astronomy, The University of Edinburgh - Edinburgh, Scotland EH9 3JZ, UK
}

\abstract{
We study colloidal particle dynamics of a model glass system using confocal and fluorescence microscopy as the sample evolves from a hard-sphere glass to a liquid with attractive interparticle interactions.  The transition from hard-sphere glass to attractive liquid is induced by short-range depletion forces.  The development of liquid-like structure is indicated by particle dynamics.   We identify particles which exhibit substantial motional events and characterize the transition using the properties of these motional events.  As samples enter the attractive liquid region, particle speed during these motional events increases by about one order of magnitude, and the particles move more cooperatively.   Interestingly, colloidal particles in the attractive liquid phase do not exhibit significantly larger displacements than particles in the hard-sphere glass.
}

\begin{document}
\maketitle

\section{Introduction}
Theory, simulation, and experiment have demonstrated that a colloidal system can be driven from a hard-sphere glass to an attractive glass by increasing short-range attractions between colloidal particles \cite{pham2002mgs,eckert2002reg,pham2004ghs,kaufman2006dir,simeonova2006doc,bergenholtz1999ntc,dawson2000hog,puertas2002css,puertas2004dhc}.  In colloidal suspensions this effect is typically realized by adding nonadsorbing polymers to the colloidal suspension.  Depletion forces \cite{asakura1954,illet1995,yodh2000abh}, induced in this way, cause the particles to move closer to one another, and the system exhibits a transition from a hard-sphere glass to an attractive liquid \cite{pham2002mgs,eckert2002reg,pham2004ghs,kaufman2006dir,simeonova2006doc}. Increasing the polymer concentration even further causes the system to enter an attractive glass phase \cite{pham2002mgs,eckert2002reg,pham2004ghs,kaufman2006dir,simeonova2006doc}.

Calculations and molecular dynamics simulations \cite{bergenholtz1999ntc,dawson2000hog,puertas2002css,puertas2004dhc} suggest that reentrance to the glass phase is due to the existence of two qualitatively different glassy states.  In hard-sphere colloidal suspensions the system enters a glass phase through a caging mechanism: as the volume fraction $\phi$ is increased, particles are increasingly trapped by their neighbors, until a critical volume fraction $\phi_{g}\sim0.58$ is reached; then caging becomes effectively permanent, stopping long-range particle motion. In attractive glasses, the attractive part of the potential causes particles to move closer to one another and eventually binds them at contacts.  In these glasses, structural arrest is due to bonding.  Thus, it is believed that the two types of glasses should have different structural and dynamical properties.

Despite numerous theoretical and experimental studies of the reentrant glass transition in colloidal suspensions, to our knowledge, only a couple of investigations have employed direct microscopic imaging to study the mechanism of this process \cite{kaufman2006dir,simeonova2006doc}.  Notably, Kaufman and Weitz \cite{kaufman2006dir} extracted qualitative information from microscopic images about particle motion magnitude and observed that particles in repulsive glasses exhibit cage rattling and escape, while in attractive glasses they exhibit large displacements upon cage escape.  In a different vein, Simeonova et. al. \cite{simeonova2006doc} reported that melting of the hard-sphere glass is accompanied by significant changes in the particle displacement distributions and their moments.  

In this Letter, we study this system class as it is brought from a hard-sphere glass into the attractive liquid region using a qualitatively different set of microscopic parameters.  Confocal microscopy experiments reveal particles that exhibit motional ``events'' wherein they move significantly farther than the thermal fluctuations within their cages.  We characterize the transition using the properties of these motional events.  Interestingly, the average displacement of particles that exhibit motional events increases only slightly as the system is brought from a hard-sphere glass to the attractive liquid.  However, the average event duration, expressed in units of Brownian time, decreases by more than an order of magnitude under the same conditions.  Thus, effective event motional speed increases with increasing interparticle attractions by almost an order of magnitude. Moreover, as the polymer-concentration-induced attractions increase, the number of particles exhibiting motional events increases by an order of magnitude, and the cluster size of particles exhibiting motional events also increases, i.e. the event motion is correlated over longer length scales.

\section{Experimental}
The particles used in this study were poly-(methylmethacrylate) (PMMA) spheres, sterically stabilized by a thin layer of poly-12-hydroxystearic acid (radius $a = 1.1$ $\mu m$, polydispersity of $\sim 5 \%$) and dyed with rhodamine.  The PMMA particles were suspended in a mixture of cyclohexylbromide/\emph{cis}- and \emph{trans}-decalin which nearly matches the density and the index of refraction of the PMMA particles to the solvent. Coulombic interactions due to surface charges on the colloidal particles were screened by adding 2 mM of tetra-butyl-ammonium chloride \cite{yethiraj2003cms}.

To induce the depletion attraction between PMMA particles, linear polystyrene polymer ($M_w = 7.5\times10^6$ Da; radius of gyration $r_g \approx 106$ nm) was added to the particle suspension.  Specifically, a series of samples with polymer concentrations, $c_p$, varying from $0-1.8$ mg/ml was   prepared using the following method. For each suspension, a colloidal sample was centrifuged to the random closed-packed volume fraction (RCP).  The sample was subsequently diluted by a mixture of the density matched liquid and polymer, yielding a sample particle volume fraction of $\phi=0.60$ (RCP in our samples was measured to be 0.66) and the desired polymer concentration.  After 24 hours of homogenization by mixing and tumbling, the colloidal suspension was loaded into a glass microscopy cell along with a small piece of magnetic wire to be used later for reinitiating the sample by stirring.  The ratio of the polymer radius of gyration to colloidal particle radius is 0.09.  

We used fluorescent and confocal microscopy to capture 2D image slices in 3D samples with a time resolution of 6 s over a time period of 3 hours.  Measurements began 10 minutes after stirring, insuring that flows within the sample had time to subside \cite{courtland2003dva}, and measurements were taken at least 35 $\mu m$ away from the cover slip surface to minimize wall effects. The position of each particle within the optical plane was obtained using standard particle tracking techniques \cite{crocker1996mdv}.

A side effect of adding polymer to induce the depletion attraction is to increase the solution viscosity \cite{pham2004ghs,kaufman2006dir,simeonova2006doc}.  As a result, at higher polymer concentrations colloidal particles diffuse more slowly. To facilitate comparisons of particle dynamics between samples with different polymer concentration and thus different viscosity, we scale the experimental time for each sample by the time it would take an isolated particle to diffuse its radius in a suspension with the same polymer concentration.  Therefore, we analyze particle dynamics in units of Brownian time $t_{B}=\frac{a^{2}}{D_{0}(c_p)}$, where $D_{0}(c_p)$ is the diffusion constant for an isolated particle in the solvent at the polymer concentration $c_p$ \cite{kaufman2006dir}.

\section{Results and Discussion}
\begin{figure}
\onefigure[scale=0.35]{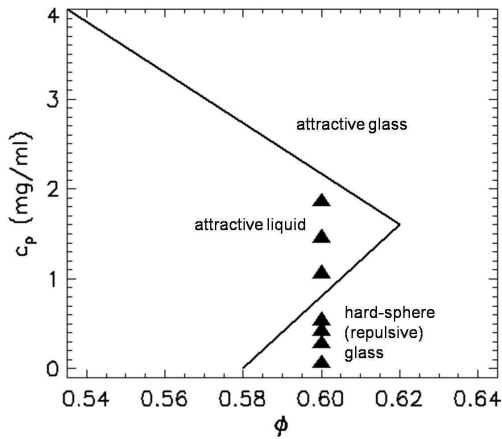}
\caption{Reentrant phase diagram with repulsive and attractive glass lines.  We plot polymer concentration $c_p$ vs. hard-sphere volume fraction. Solid triangles indicate samples studied here and follow a transition from a hard-sphere glass to the attractive liquid region.}
\label{phasediag}
\end{figure}

Figure \ref{phasediag} presents a phase diagram of the reentrant glass transition.  Solid triangles in Fig. \ref{phasediag} correspond to the samples studied, starting from a hard-sphere glass ($c_p = 0$ mg/ml) and ending in the attractive liquid region.  The solid lines are schematic, qualitatively indicating the glass transition boundaries.  We estimated the lower boundary using sample crystallization (e.g., via the bond orientational order parameter $\Psi_6$).  The upper boundary is a conjecture.  Samples in the attractive liquid region showed significant crystallization compared to samples in the hard-sphere glass region.

Evidence for the reentrant glass transition has been derived in the past \cite{pham2002mgs,eckert2002reg,pham2004ghs,kaufman2006dir,simeonova2006doc}.  Here it is apparent in the plot of $D(c_p)/D_0(c_p)$ vs. $c_p$, where $D(c_p)$ is the long time diffusion constant of the particles in suspension at high $\phi$ with polymer concentration $c_p$.  Indeed, $D(c_p)/D_0(c_p)$ changes by one to two orders of magnitude as the system evolves from a hard-sphere glass ($c_p = 0$ mg/ml) into the attractive liquid region ($c_p \sim$ 2.0 mg/ml).

\begin{figure}
\onefigure[scale=0.3]{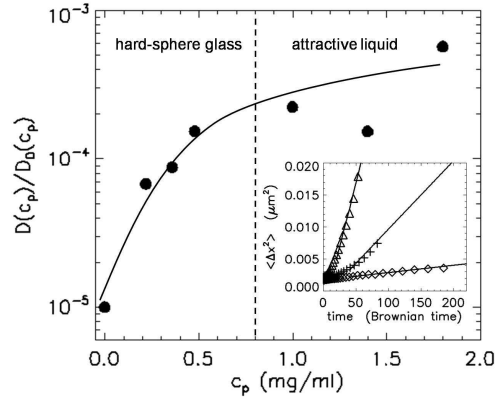}
\caption{Long time diffusion constant $D(c_p)$ divided by bare diffusion constant $D_0(c_p)$
as a function of polymer concentration $c_p$.  Solid line is to guide the eye.  Dashed line indicates approximate phase boundary
between hard-sphere glass and attractive liquid.  Inset shows mean square displacements vs. Brownian time for representative polymer concentrations of 0 mg/ml ($\diamond$), 0.22 mg/ml ($+$), 1.8 mg/ml ($\triangle$).  Solid lines denote linear fits to data at long times (i.e. in the more diffusive regime).  The slopes of these lines were used to determine long time diffusion constants $D(c_p)$.
}
\label{DoverD0}
\end{figure}

\begin{figure}
\onefigure[scale=0.4]{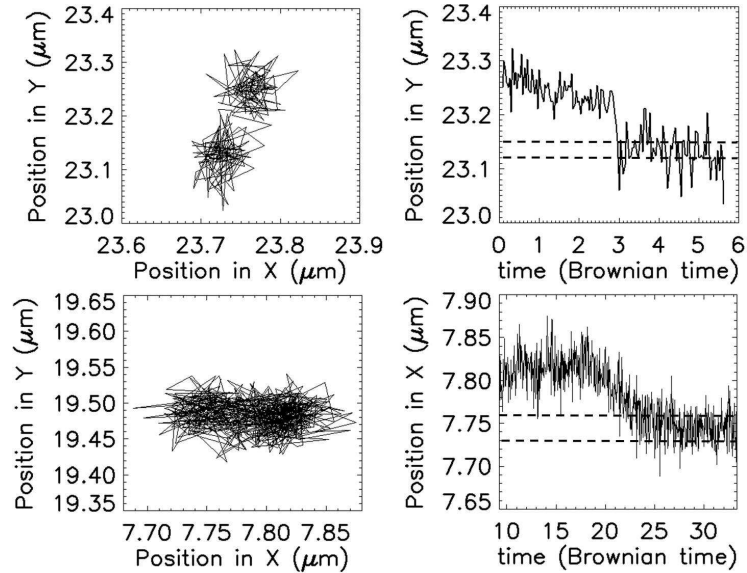}
\caption{
Spatial images of two classes of particle motional events.  Note that the time scale of motional events varies.  The top row shows a short-time scale motional event (particle jump), whereas the bottom row shows an example of more gradual motional event.  Dashed lines approximately bound a region of twice the standard deviation, $\sigma$, of the particle position.
}
\label{exampleevents}
\end{figure}

Microscopy studies of colloidal suspensions permit determination of particle positions during the entire experiment.  Thus, it is possible to quantify the behavior of the most ``active'' particles in suspension; these ``active'' particles have been of particular interest recently \cite{weeks2000tdd,kegel2000,weeks2002pcr,vollmayrlee2001dhb}.  We define particles to be active when they move significantly farther than the thermal fluctuations in their cages, following the definition of particle ``jumps'' in Ref. \cite{vollmayrlee2001dhb} (see top row Fig. \ref{exampleevents}). 

For each particle, we calculate running-five-point average position: 
$\overline{r}(t)=\frac{1}{5}\sum^{t+2}_{i=t-2}r(i)$. 
Next, we calculate the change in this average particle position, $\Delta \overline{r}$, 
during the time interval $\Delta t$: 
$\Delta \overline{r(t)} =
\overline{r(t)}-\overline{r(t-\Delta t)}$.
Finally, we compare the average displacement $\Delta \overline{r}$ with
average fluctuations ($\sigma$) of the particle during the entire time,
$T$, that the particle is tracked:
$\sigma^2=\frac{1}{N}\sum^{i=N}_{i=1}(\overline{r(t_i)^2}-\overline{r(t_i)}^2)$
where
$N$ is the total number of time steps (see Fig. \ref{exampleevents}).
If
$\Delta \overline{r(t)} > \sqrt{20}\sigma$, then
we say that at time $t$ the particle exhibited a \emph{motional event} of duration $\Delta t$ (for further details see Ref.\cite{vollmayrlee2001dhb}). Typically, $r(t)$ is constant with fluctuations before and after a motional event.  To properly identify events, it is important to choose an appropriate $\Delta t$. A short $\Delta t$, e.g. comparable to one Brownian unit, is useful for identifying quick events (jumps), but it is not suitable for identifying more gradual motions; we observe a wide range of motional event durations, including ones that last more than several Brownian time units (bottom row Fig. \ref{exampleevents}). Therefore, to identify particles that change their positions over relatively longer times, we perform the above calculation for a range of  $\Delta t$'s.  Furthermore, we only include particles which exhibit a complete motional event, i.e. an event that began and ended during the time of the experiment.

\begin{figure}
\onefigure[scale=0.4]{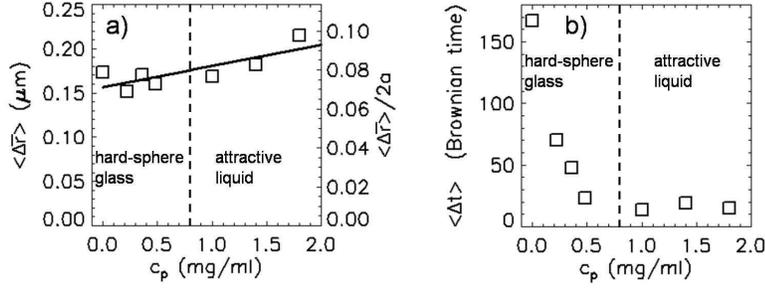}
\caption{a) Average displacement $\langle \Delta \overline{r} \rangle$ of particles exhibiting events vs. polymer 
concentration.  Right axis denotes $\langle \Delta \overline{r} \rangle$ scaled by particle diameter $2a$. Solid line is 
a least-squares fit to the data with slope 0.24.  b) Average duration $\langle \Delta t \rangle$ of motional 
events vs. polymer concentration.  Event duration is in units of Brownian time.  Dashed lines indicate approximate phase boundary
between hard-sphere glass and attractive liquid.
}
\label{drdt}
\end{figure}

We first focus attention on the distribution of position displacements, $\Delta \overline{r}$, for motional events.  Average particle displacement $\langle \Delta \overline{r} \rangle$ during an event vs. polymer concentration is shown in Figure \ref{drdt}a.  Interestingly, as the polymer concentration increases, particles exhibiting motional events travel further by only about 0.05 $\mu m$, about one tenth of the particle diameter.  This small displacement is on the order of the cage Brownian fluctuations and tracking uncertainty.  The observation is somewhat counterintuitive, since one might expect that with increasing polymer concentration, the number of colloidal particles that become stuck to each other increases, thus creating more free space for other particles to jump \cite{simeonova2006doc}.  Our data do not appear to support such a conjecture.

Similarly, we analyzed distributions of event durations.  Figure \ref{drdt}b presents the average event duration (in Brownian units) vs. polymer concentration.  Average event duration decreases from about 170 Brownian units for $c_p = 0$ mg/ml to about 15 Brownian units for $c_p = 0.8$ mg/ml.  For polymer concentrations in the attractive liquid region, the average event time saturates at about 15 Brownian units.  Thus, as the polymer concentration increases, particles that exhibit motional events do so in a shorter time until the attractive liquid region is reached, wherein all motional events take approximately the same time.

\begin{figure}
\onefigure[scale=0.3]{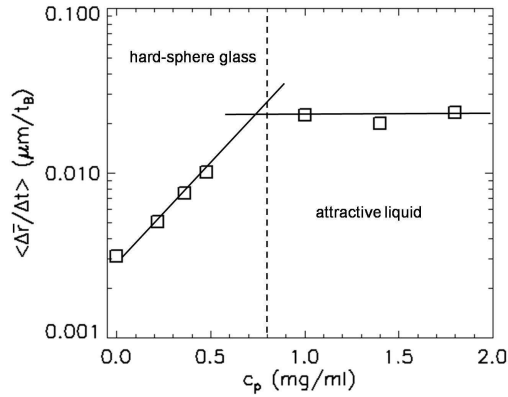}
\caption{Average event speed $\langle \Delta \overline{r}/\Delta t \rangle$ in units of $\mu m/t_B$ vs. polymer concentration.  
Solid lines are guide to eyes.  Dashed line indicates approximate phase boundary between hard-sphere glass and attractive liquid.
}
\label{eventspeed}
\end{figure}

From event displacement and duration information, we calculate particle motional event speed, $\langle \Delta \overline{r}/\Delta t \rangle$, and on Fig. \ref{eventspeed} we plot average particle event speed vs. polymer concentration.  Particles experiencing motional events move faster with increasing polymer concentration.  For the samples in the vicinity of the attractive liquid region, the event speed changes by almost an order of magnitude with respect to the event speed in the hard-sphere sample.  Then, for polymer concentrations farther into the attractive liquid region, the event speed saturates.  

\begin{figure}
\onefigure[scale=0.35]{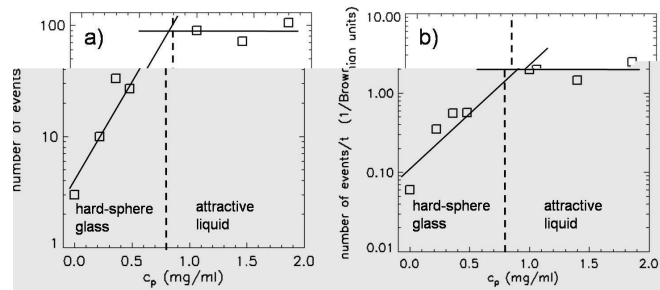}
\caption{a) Number of motional events vs. polymer concentration calculated from data of the
same duration in Brownian time units.  b) Rate of motional events  vs. polymer concentration.  
Solid lines are guide to eyes.  Dashed lines indicate approximate phase boundary
between hard-sphere glass and attractive liquid.}
\label{eventsall}
\end{figure}

We next consider the raw number of particle events as a function of polymer concentration as shown in Figure \ref{eventsall}a.  The number of particle events initially increases from about 3 to almost 100 and then saturates for polymer concentrations in the attractive liquid region.  We might expect that as the attractive glass phase is approached, the number of motional events would decrease.  However, such a plot (Fig. \ref{eventsall}a) does not account for the increase of solvent viscosity with increasing polymer concentration.  Thus, we calculate event rate by scaling the number of events by the length of the data in units of Brownian time $t_B$ (Fig. \ref{eventsall}b).   The ``viscosity normalized'' event rate increases with polymer concentration by more than an order of magnitude until polymer concentrations of about 1 mg/ml are reached.  Therefore, the number of particles that exhibit motional events, and hence are responsible for the relaxation in the samples, increases significantly with polymer concentration as the system fluidizes.

\begin{figure}
\onefigure[scale=0.35]{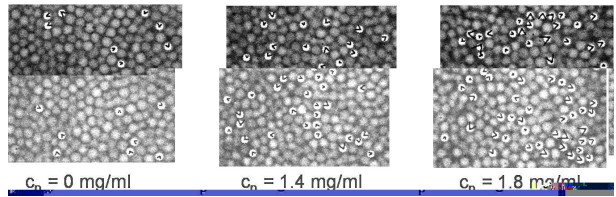}
\caption{Microscopy images for polymer concentrations of 0 mg/ml, 1.4 mg/ml, and 1.8 mg/ml.
White dots are plotted over the particles that are exhibiting events.  Arrows on the white dots indicate the direction of
motion.}
\label{flowhistrogramsandpics}
\end{figure}

To analyze the collective behaviors of particles exhibiting motional events further, we examine the spatial distributions of particles that exhibit motional events.  Figure \ref{flowhistrogramsandpics} shows representative microscopy snapshots for three polymer concentrations.  White dots are plotted over particles that exhibit motional events with arrows indicating the direction of the motion.  As the polymer concentration increases, a particle that exhibited a motional event has,  on average, more neighbors that are also moving significantly.  Thus, with increasing polymer concentration more particles are moving cooperatively.  

\begin{figure}
\onefigure[scale=0.35]{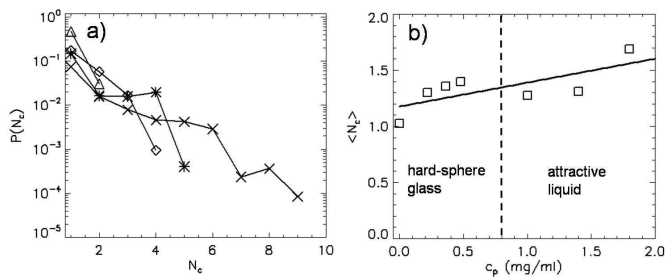}
\caption{a) Probability distribution of cluster size (number of particles $N_c$) for the following representative polymer concentrations: 0 mg/ml ($\triangle$), 0.36 mg/ml ($\diamond$), 0.48 mg/ml ($\ast$), 1.8 mg/ml ($\times$).  b) Average cluster size $\langle N_c \rangle$ at different polymer concentrations $c_p$.  Solid line is a least-squares fit to the data linear with slope 0.21.  Dashed line indicates approximate phase boundary between hard-sphere glass and attractive liquid.
}
\label{clusters}
\end{figure}

To look for spatial correlations of the particles that exhibited a motional event, we analyze the nearest neighbor connectivity and thus identify clusters of connected particles that exhibit a motional event.  In Fig. \ref{clusters} we plot the frequency distribution of cluster size, $P(N_c)$, vs. number of particles in a cluster, $N_c$, for representative polymer concentrations.  For low polymer concentrations, particles have a tendency to move in small clusters.  As the polymer concentration increases, the particles move in increasingly bigger clusters.  At the highest polymer concentration studied here, $c_p=1.8$ mg/ml, we observed clusters composed of as many as ten particles.  Thus, as the attractive liquid region is approached, structural relaxation occurs because of the motion of small numbers of large cooperative clusters of particles that exhibit motional events, rather than mostly solitary particles, as we observe in hard-sphere glass.  This effect is also indicated by average cluster size, $N_c$.  The average cluster size increases by almost a factor of two with polymer concentration as shown in Figure \ref{clusters}b.  However, the size distribution of the clusters of particles that exhibited a motional event is likely even broader than presented here since clusters may extend beyond the viewing area of x-y focal plane.

In summary, we have studied a colloidal system with short-range attractive potential in the reentrant region using primarily the properties of ``motional events''.  We observe the transition from a hard-sphere arrested phase to a liquid-like phase.  This transition is characterized by increase in: $D(c_p)/D_0(c_p)$, event speed, and the event rate of moving particles.  Interestingly, particles exhibiting a motional event do not move longer distances at higher polymer concentration, but they do move faster (i.e. in Browniant time units).  The transition to the reentrant region is also characterized by a growing number of particles that experience motional events.  Moreover, the particles experiencing motional events are increasingly spatially correlated with increasing attraction.  The particles move in clusters, and the distribution of the cluster size becomes broader and shifts to larger average values with increasing interparticle attraction.  

Future microscopy studies should include exploration of re-entrance into the attractive glass region and possibly the influence of particle-to-polymer size ratio on the system dynamics \cite{ren1993}.  Also, since our studies provide only 2-dimensional information, 3-dimensional studies should shed more light on the cluster size and distance traveled by the particles exhibiting motional events.  Finally, similar systematic studies of particle dynamics in colloidal suspensions with short-range attractions along the low volume fraction extension of the attractive glass phase line may lead to a unified description of glasses and gels.

\acknowledgments
We thank Paul J. Angiolillo, Katharina Vollmayr-Lee, Eric R. Weeks, Peter Yunker, and Zexin Zhang for helpful discussions, and Paul J. Angiolillo for comments on the manuscript.  PH acknowledges financial support from an award from Research Corporation and Sigma Xi Scientific Research Society, SJU Chapter.  AGY acknowledges partial support from the NSF DMR-052002 (MRSEC) and DMR-0804881.

\end{document}